\documentclass[prb,twocolumn,amssymb,amsfonts,superscriptaddress,showpacs]{revtex4}
\usepackage{epsfig}
\usepackage{amsmath}
\usepackage{bm}

\begin{document}

\title{\bf{Indirect coupling between spins in semiconductor quantum dots}}

\author{G.~Ramon}
\email{ramon@bloch.nrl.navy.mil}
\affiliation{Naval Research
Laboratory, Washington DC 20375-5320}
\author{Y.~Lyanda-Geller}
\altaffiliation[Current address: ]{Department of Physics, Purdue
University, West Lafayette, Indiana 47907.}
\affiliation{Naval
Research Laboratory, Washington DC 20375-5320}
\author{T.~L.~Reinecke}
\affiliation{Naval Research Laboratory, Washington DC 20375-5320}
\author{L.~J.~Sham}
\affiliation{Department of Physics, University of California San
Diego, La Jolla, California 92093-0319}

\begin{abstract}
The optically induced indirect exchange interaction between spins
in two quantum dots is investigated theoretically. We present a
microscopic formulation of the interaction between the localized
spin and the itinerant carriers including the effects of
correlation, using a set of canonical transformations. Correlation
effects are found to be of comparable magnitude as the direct
exchange. We give quantitative results for realistic quantum dot
geometries and find the largest couplings for one dimensional
systems.
\end{abstract}


\maketitle

Control of spins in semiconductors has been intensively
investigated in recent years due to its potential for applications
in spintronics and quantum computation \cite{Aws}. Coherent
coupling between localized spins is particularly sought after
because it is a key requirement in any proposal for spin-based
implementation of quantum computation. Several coupling mechanisms
have been proposed to construct quantum gates between spins in
quantum dots (QDs). These include direct wavefunction overlap
using electric gates where the interdot separation is small
\cite{LosDiv} and exchange of a cavity photon mode between spins
in QDs for a large interdot separation \cite{Ima}.

Recently Piermarocchi {\it et al.} proposed to use an indirect
mechanism to couple the two QD spins at intermediate interdot
separations \cite{Pier}. Here the interaction is mediated by
virtual delocalized carrier excitations in the host material. The
virtual excitations are driven by an interband off-resonance laser
that provides optical control over the interaction and serves to
reduce the bandgap energy, thus extending the interaction range.
The proposed scheme has the advantages of ultrafast optical
control and long spin coherence times due to the virtual nature of
the excitations. Combined with the proposal to use Raman optical
transitions via intermediate trion states for single qubit
operations \cite{1qbit}, this mechanism provides the necessary set
of universal gates for quantum computing. This optically induced
indirect spin exchange is a variant of several analogous
mechanisms. These include the RKKY interaction in metals
\cite{RKKY} and Bloembergen-Rowland coupling in direct-gap
semiconductors \cite{BR}, the superexchange mediated by two holes
in diluted magnetic semiconductors (DMS) \cite{Fur}, the magnetic
exchange mediated by bound correlated states (excitons)
\cite{Melo,Pier2}, and ferromagnetism induced by virtual Mn
acceptor level-valence band transitions in DMS materials
\cite{Lit}.

A key ingredient in all of these indirect spin coupling mechanisms
is the exchange interaction of a localized spin with the mediating
itinerant excitation. The purpose of the present work is to
introduce a microscopic formulation that provides a quantitative
description of this exchange interaction by taking into account
the effects of hybridization of continuum and dot states and the
double occupancy in the dot.

For the case of the optically induced indirect interaction between
spins in quantum dots, the spin-spin coupling is obtained by
considering the self-energy correction in the continuum electron
propagator due to its Coulomb interaction with each of the
localized spins within second order perturbation theory
\cite{Pier}. The result is a Heisenberg Hamiltonian involving the
localized spins, with an effective positive exchange constant that
is given by \cite{Pier}
\begin{equation}
J_{12}(R)=\frac{|\Omega |^2}{16} \int \frac{d^d k d^d k'}{(2
\pi)^{2 d}} \frac{|J({\bf k},{\bf k}')|^2 {\rm e}^{-{\rm i} ({\bf
k}-{\bf k}') \cdot {\bf R}}}{ \left( \delta+\frac{k^2}{2\mu}
\right)^2 \left( \delta +\frac{k^2}{2m_{\rm
h}}+\frac{k'^2}{2m_{\rm e}} \right)} \label{J12}
\end{equation}
where $R$ is the distance between the dot centers, $\delta$ is the
detuning of the laser with respect to the electron-hole continuum,
$\Omega$ is the Rabi energy for the light coupling to the
electron-hole pair and $\mu$ is its reduced mass. $J({\bf k},{\bf
k}')$ is the exchange interaction between the electron spin in the
quantum dot and the itinerant electron.

To calculate $J({\bf k},{\bf k}')$ we consider a Hamiltonian that
includes the kinetic energy, the dot potential relative to the
host material and the electron-electron Coulomb interactions:
\begin{equation}
{\cal H}={\cal H}_0+{\cal H}_M+ {\cal H}_1 \label{Ham}
\end{equation}
where
\begin{subequations}
\begin{eqnarray}
\mathcal{H}_0 \!\!&=&\!\!\sum_{\sigma} E_d n_{\sigma}
+\sum_{\mathbf{k},\sigma}E_{k} c_{\mathbf{k}\sigma }^{\dagger
}c_{\mathbf{k}\sigma }+U n_{\uparrow }n_{\downarrow}
\label{H0} \\
\mathcal{H}_M \!\!&=&\!\! \sum_{\mathbf{k},\sigma } \left[ V_k
c_{\mathbf{k}\sigma} ^{\dagger }c_{d\sigma }+T_k
c_{\mathbf{k}\sigma }^{\dagger }c_{d \sigma }n_{-\sigma} +h.c.
\right] \label{HM} \\
\mathcal{H}_{1} \!\!&=&\!\!
\sum_{\stackrel{\mathbf{k}\mathbf{k}^{\prime}}{\sigma \sigma
^{\prime }}}C^{\rm dir}_{{\bf k}{\bf k}'} c_{\mathbf{k} \sigma
}^{\dagger } c_{\mathbf{k}^{\prime }\sigma
}n_{\sigma ^{\prime }}+\sum_{%
\mathbf{k}\mathbf{k}^{\prime }\sigma } C^{\rm ex}_{\mathbf{k}
\mathbf{k}'} c_{\mathbf{k}\sigma }^{\dagger } c_{d\sigma }
c_{d-\sigma }^{\dagger } c_{\mathbf{k}^{\prime}-\sigma } \nonumber
\\ && +\sum_{\mathbf{k}\mathbf{k}^{\prime},\sigma} C^{\rm mix}_{{\bf
k}{\bf k}'} c_{\mathbf{k} \sigma }^{\dagger } c_{d\sigma }c_{{\bf
k}' -\sigma}^{\dagger }c_{d -\sigma } +h.c.  \label{H1}
\end{eqnarray}
\end{subequations}
Here $c^\dagger_{d \sigma}$ ($c^\dagger_{{\bf k} \sigma}$) is the
creation operator for a localized (itinerant) electron,
$n_{\sigma}=c^{\dagger}_{d \sigma} c_{d \sigma}$, and the last
term in Eq.~\eqref{H0} is the on-site Coulomb repulsion. ${\cal
H}_M$ represents the hybridization of the localized and itinerant
electrons, where we include a population dependent hybridization
(2nd term in Eq.~\eqref{HM}), which was absent in previous works
on coupling of localized and itinerant spins. We show below that
this latter term makes an important contribution to the spin
exchange interaction. $\mathcal{H}_1$ contains the
spin-independent and spin exchange Coulomb scattering; the latter
gives rise to the Heitler-London exchange contribution \cite{Ash}.
The last term in Eq.~\eqref{H1} describes the effect of localized
and continuum state mixing. We have neglected scattering processes
between carriers in the continuum since the corresponding effects
are not relevant to the problem we wish to solve. $V_k=\int d{\bf
r} V_{\rm dot} \varphi_{\bf k}^*({\bf r}) \varphi_d({\bf r})$ is
the tunnelling amplitude, where $\varphi_{\mathbf{k}} (r)$
[$\varphi_d (r)$] is the itinerant (localized) electron wave
function, also used to calculate he various Coulomb integrals in
Eq.~\eqref{Ham}.

We aim at bringing the Hamiltonian ${\cal H}'={\cal H}_0+{\cal
H}_M$ to a form similar to that of an s-d spin exchange
Hamiltonian by applying a canonical transformation
\begin{equation}
{\cal \bar{H}}'={\rm e}^{S}{\cal H}'{\rm e}^{-S}.
\end{equation}
The unitary operator $S$ is constructed to eliminate ${\cal H}_M$
to first order by requiring ${\cal H}_M=-[S,{\cal H}_0]$ and is
given by
\begin{equation}
S=\sum_{{\bf k} \sigma} \left[
\beta_k+(\alpha_k-\beta_k)n_{-\sigma}\right] c_{d \sigma}^\dag
c_{{\bf k} \sigma} - h.c. \label{S}
\end{equation}
where
\begin{equation}
\alpha_k=\frac{V_k+T_k}{U+E_d-E_k} \ \ ; \ \
\beta_k=\frac{V_k}{E_d-E_k}. \label{ab}
\end{equation}
This is a generalized form of the Schrieffer-Wolff transformation,
which was first used to establish the connection between the
Anderson and Kondo models \cite{SW}. It produces a contribution to
the spin exchange arising from the correlation and hybridization
terms in ${\cal H}'$, which is given to first order by
\begin{equation}
J^{(1)}(k,k') = \frac{1}{2} \left[ \beta_k V^*_{k'}- \alpha_k
(V_{k'}+T_{k'})^* \right]+\left[ k\leftrightarrow k'\right]^*.
\label{exchange}
\end{equation}
This contribution vanishes when correlation effects are neglected
($U$, $T_k \rightarrow 0$). We find that the first order result,
given in Eq.~\eqref{exchange}, is inadequate because it requires
that ${\cal H}_M$ would be a small perturbation to ${\cal H}_0$,
which is not the case generally. It is therefore necessary to sum
up the infinite series in the transformed Hamiltonian
\begin{equation}
\bar{{\cal H}}'={\cal H}_0+\sum_{n=1}^{\infty}
\left(\frac{1}{n!}-\frac{1}{(n+1)!} \right) [S,{\cal H}_M]_n,
\label{Hbar}
\end{equation}
where $[S,{\cal H}_M]_n=[S,[S, ...,[S,{\cal H}_M]...]]$. To this
end we use a method suggested by Chan and Gul\'{a}csi \cite{CG}
but employ a different strategy to solve the problem. The first
term in the series in Eq.~\eqref{Hbar} is
\begin{eqnarray}
[S\!\!\!&\!,\!\!\!&\!{\cal H}_M]_1 = \sum_{{\bf
k}{\bf k}',\sigma} \left[ J_1(k,k') \left( c_{{\bf k}
\sigma}^{\dagger}c_{d \sigma} c_{d -\sigma}^{\dagger}c_{{\bf k}'
-\sigma} + \right. \right.
\nonumber \\
&&\left. \left. n_{-\sigma} c_{{\bf k} \sigma}^{\dagger} c_{{\bf
k}' \sigma} \right) +P_1(k,k') \left( c_{{\bf k} \sigma}^{\dagger}
c_{d \sigma} c_{{\bf k}' -\sigma}^{\dagger}
c_{d -\sigma} + h.c. \right) \right. \nonumber \\
&&\left. -K_1(k,k') c_{{\bf k} \sigma}^{\dagger} c_{{\bf k}'
\sigma} \right] +\sum_\sigma \left[ G_1 n_{\sigma} +I_1 n_{\sigma}
n_{-\sigma} \right]\label{O1}
\end{eqnarray}
where $J_1(k,k')=2 J^{(1)}(k,k')$ is given in
Eq.~\eqref{exchange}, and the rest of the coefficients in
Eq.~(\ref{O1}) are $P_1(k,k')= \frac{1}{2} \left[ \alpha_k
V^*_{k'}-\beta_k(V_{k'}+T_{k'})^*\right] +[k \leftrightarrow
k']^*$, $K_1(k,k')= \beta_k V^*_{k'}+\beta^*_{k'} V_{k}$, $G_1=
2\sum_{{\bf k}} \beta_k V^*_k$, and $I_1= 2\sum_{{\bf k}}
\left[\alpha_k (V_k+T_k)^* -\beta_k V^*_k\right]$. Calculating the
second order term we find that it has the same form of ${\cal
H}_M$ apart from higher order correlation terms. We estimate the
magnitude of these continuum scattering terms by neglecting
off-diagonal contributions and placing lower and upper bounds on
the occupation numbers. This procedure brings all higher odd
orders in the series to the form of Eq.~\eqref{O1}, and we are
able to sum the series by solving the following set of recursion
relations for the several coefficients
\begin{eqnarray}
J_{m+1}(k\!\!\!&\!,\!\!\!&\!k') = 2G_m \left(\alpha_k
\alpha_{k'}^*- \beta_k \beta_{k'}^* \right) +4I_m \alpha_k
\alpha_{k'}^* - \nonumber \\
&& \!\!\!\!\!\!\!\!\!\!\!\! \sum_{{\bf k}''} \left\{
\left[2J_m(k,k'') \alpha_{k''} \alpha^*_{k'}+2P_m(k,k'')
\beta_{k''} \alpha_{k'}^* - \right. \right. \nonumber \\
&& \!\!\!\!\!\!\!\!\!\!\!\!  \left. \left. K_m(k,k'')\left(
\alpha_{k''} \alpha_{k'}^* -\beta_{k''} \beta_{k'}^* \right)
\right] +[k\leftrightarrow k']^* \right\} \nonumber \\
P_{m+1}(k\!\!\!&\!,\!\!\!&\!k')= I_m \left( \beta_k
\alpha_{k'}^*+\alpha_k \beta_{k'}^* \right)- \nonumber
\\ && \!\!\!\!\!\!\!\!\!\!\!\! \sum_{{\bf k}''} \left\{ \left[J_m(k,k'') \alpha_{k''}
\beta_{k'}^*+P_m(k,k'')
\beta_{k''} \beta_{k'}^* - \right. \right. \nonumber \\
&&\!\!\!\!\!\!\!\!\!\!\!\!  \left. \left. \frac{1}{2} K_m(k,k'')
\left(\alpha_{k''} \beta_{k'}^*-\beta_{k''} \alpha_{k'}^* \right)
\right]+[k\leftrightarrow k']^* \right\}
\nonumber \\
K_{m+1}(k\!\!\!&\!,\!\!\!&\!k')= -2G_m \beta_k \beta_{k'}^* -
\nonumber \\ && \!\!\!\!\!\!\!\!\!\!\!\! \sum_{{\bf k}''} \left[
K_{m}(k,k'') \beta_{k''}\beta_{k'}^*+(k\leftrightarrow k')^*
\right] \label{rec} \\
G_{m+1}&& \!\!\!\!\!\!=-2G_m\sum_{{\bf k}} |\beta_k|^2-2\sum_{{\bf
k},{\bf k}'}K_m(k,k')
\beta_k \beta_{k'}^* \nonumber \\
I_{m+1}&& \!\!\!\!\!\!=-2\sum_{{\bf k}} \left[ G_m \left(
|\alpha_k|^2-|\beta_k|^2 \right)+2I_m |\alpha_k|^2
\right]+\nonumber
\\ && \!\!\!\!\!\!\!\!\!\!\!\! 2\sum_{{\bf k},{\bf k}'} \left[
P_m(k,k') \left( \beta_k \alpha_{k'}^*+\alpha_k \beta_{k'}^*
\right) + \right. \nonumber
\\ && \!\!\!\!\!\!\!\!\!\!\!\! \left. 2J_m(k,k') \alpha_k \alpha_{k'}^* -
K_m(k,k')\left( \alpha_k \alpha_{k'}^*-\beta_k
\beta_{k'}^* \right) \right] \nonumber
\end{eqnarray}
Equations \eqref{rec} are obtained from the lower bound in the
higher order contributions, and a second set of equations is
obtained for the upper bound case.

The exchange contribution is obtained from the odd orders of the
series, which also contain additional terms that renormalize the
original Hamiltonian \eqref{H0}. The even orders also are summed
up and renormalize the hybridization Hamiltonian \eqref{HM}.
Figure 1a shows the result of the series summation for the
diagonal part of the exchange, $J(k,k)$. Since it differs
appreciably from the first order result of Eq.~\eqref{exchange},
the residual hybridization in the even orders need not be small,
as seen in figure 1b. Thus we need to perform a second canonical
transformation that is defined by applying Eqs.~(\ref{S}-\ref{ab})
to our renormalized Hamiltonian. This second transformation
eliminates the next order in the hybridization terms and further
corrects the resulting exchange contribution. The process is
reiterated until we fully eliminate the hybridization part of the
Hamiltonian, as shown in figure 1b.

This procedure of applying a set of nested Schrieffer-Wolff
transformations is essential to obtain quantitative results for
the kinetic exchange interaction, which is the one that results
from the hybridization terms in Eq.~\eqref{HM}. As seen in Figure
1a, $J({\bf k},{\bf k}')$ is ferromagnetic after one
transformation, which would differ from other results for this
kinetic exchange contribution, e.g., in a renormalization group
approach \cite{Kri}.
Only after a set of transformations (typically 10-20) are the
renormalized hybridization terms eliminated and the
antiferromagnetic nature of the interaction is restored, albeit
with a modified magnitude compared to the first order result. The
results calculated with the lower and upper bounds discussed above
are remarkably close to one another. We have verified that they
coincide within $10\%$ for a wide range of geometries and dot
potentials. In order to address the possibility that the
off-diagonal contributions from the continuum scattering terms
might alter our results, we have estimated them from limits of the
off diagonal density factors. The results lie between the lower
and upper bounds given in figure 1a; thus we believe that our
summation represents the complete Schrieffer-Wolff transformation
with a good accuracy.
\begin{figure}[!tb]
\epsfxsize=0.77\columnwidth \centerline{\epsffile{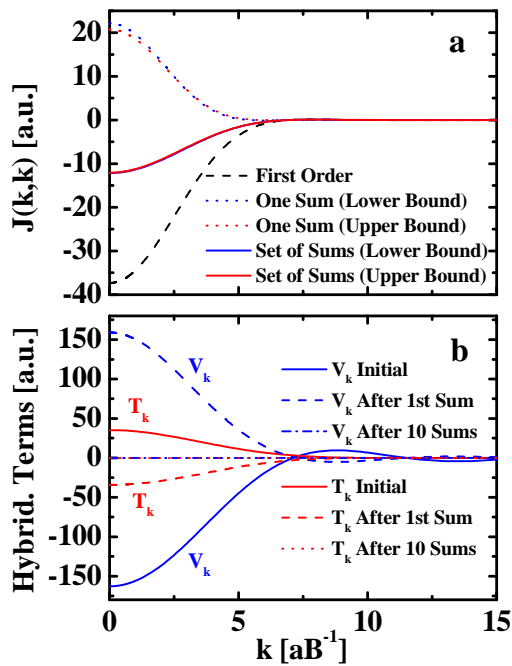}}
\vspace*{0.cm} \caption{(color online) (a) Diagonal matrix
elements of the spin exchange interaction between localized and
itinerant electrons in a 2D host. The figure presents the first
order result from Eq.~\eqref{exchange} (dashed line), intermediate
results after the summation in Eq.~\eqref{Hbar} (dotted lines),
and final results obtained by performing a set of transformations
(solid lines); (b) The corresponding hybridization terms:
tunnelling amplitude, $V_k$, and population-dependent
hybridization term, $T_k$.} \label{Jkk}
\end{figure}

Since the kinetic exchange interaction that we calculate from the
transformed ${\cal H}'$ is antiferromagnetic, it competes with the
ferromagnetic exchange given by the second term in Eq.~\eqref{H1}.
Thus, an accurate evaluation of the former is important as it can
lead to an order of magnitude difference or even a change of sign
in the total spin exchange coupling between localized and
itinerant electrons. A full transformation is also valuable in the
case where $U+E_d>0$, leading to a divergence of $\alpha_k$ in
Eq.~\eqref{ab}. Here, the kinetic exchange contribution is
dominant and cannot be obtained via a perturbative approach.
This regime corresponds to dots with small size ($R_D \leq 5nm$)
and shallow potential (barrier $\leq 80meV$), which are not
typical for physical systems and are not considered here.

\begin{figure}[!tb]
\epsfxsize=0.75\columnwidth \vspace*{-0.1cm}
\centerline{\epsffile{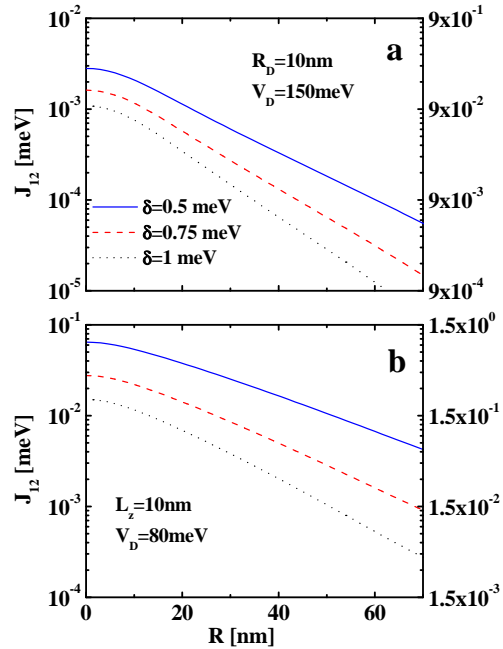}} \vspace*{0.cm} \caption{(color
online) (a) Optically induced spin-spin exchange coupling in a 2D
host vs. the distance between the centers of the dots for dot
radius $R_D=10nm$, potential height $V_e=150meV$ and several
values of laser detunings. Right axis shows the coupling values
including excitonic effects; (b) Same as a for a quasi 1D host and
cylindrical dots with $L_z =10nm$, $R_D=5nm$ and potential height
$V_e=80meV$.} \label{J12_2D_1D}
\end{figure}
In figure 2 we show the results for the spin-spin coupling
$J_{12}$ [Eq.~\eqref{J12}], incorporating all the exchange
contributions. Figure 2a shows the spin coupling for lateral
cylindrical dots in a two-dimensional quantum well. The results
for vertically stacked cylindrical dots in a quasi one-dimensional
wire are given in figure 2b. Here we used $m_e=0.07m$, $m_h=0.5m$
and $\Omega=0.1meV$. The localized electron wave function was
taken to be a combination of Bessel functions in the lateral
direction and a combination of Cosine and Exponential functions in
the $z$ direction. It is seen that the spin coupling is more than
an order of magnitude larger for the one-dimensional case than for
the two-dimensional case.

The Coulomb interaction between the intermediate virtual electrons
and holes results in an enhancement of the oscillator strength at
the optical and spin vertices due to the exciton wave functions
\cite{Pier}. We have evaluated the dominant contribution of the
electron-hole interaction to $J_{12}$. It results in an increase
of up to two orders of magnitude in the two-dimensional case and
roughly one order of magnitude in the one-dimensional case (right
axes in figure 2). Thus, the excitonic effects reduce the
difference in $J_{12}$ between the two geometries.


Figure 3 shows the dependence of the spin-spin coupling on the dot
potential and size. Larger dots give larger couplings but
necessitate larger separation to avoid overlap. The increase in
the coupling as the dot potential decreases is mainly due to the
reduction of the kinetic exchange contribution.
\begin{figure}[!tb]
\epsfxsize=0.75\columnwidth \vspace*{0.cm}
\centerline{\epsffile{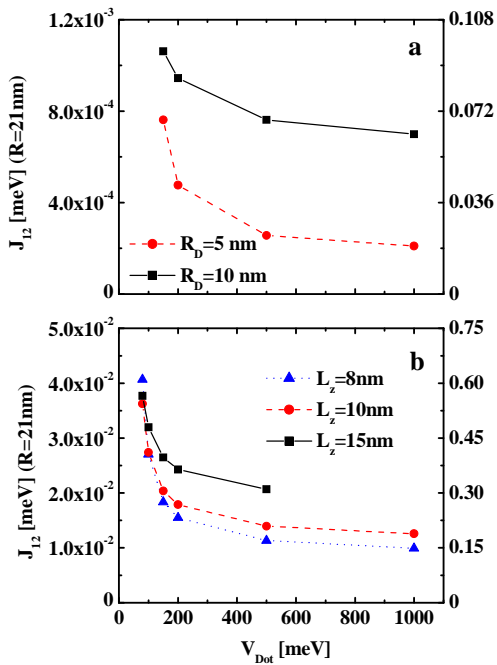}} \vspace*{0.cm} \caption{(color
online) (a) Spin-spin coupling in a 2D host at a dot separation of
$21nm$ vs. the dot potential for two dot radii. The laser detuning
is $\delta=0.5meV$. Right axis shows the coupling values after
excitonic corrections; (b) Same as a for a quasi 1D host and
cylindrical dots with $R_D=5nm$ and several dot heights.}
\label{J12_pot}
\end{figure}

A technologically viable way to increase $J_{12}$ is by using a
microcavity. This can be done by growing distributed Bragg
Reflector layers on the top and bottom of the active semiconductor
layer containing the QDs. Placing the active layer at the antinode
of the microcavity increases the electric field by orders of
magnitude, and thus increases the Rabi energy at the optical
vertices in Eq.~\eqref{J12}.

We have shown that the effect of hybridization of continuum and
dot states produces a sizable contribution to the exchange
coupling between localized and itinerant electrons. For certain
dot geometries this kinetic exchange can even lead to a change of
sign in the spin exchange interaction. A set of canonical
transformations with summations over higher order terms provides a
useful tool to evaluate the spin exchange interaction. Our
transformation of the Hamiltonian \eqref{Ham} captures the
multiple scattering processes involved in the interaction between
the localized and itinerant carriers, and it provides the first
microscopic description that accounts quantitatively for the
exchange interaction \cite{Pier-note}. Our formulation is also
applicable to other systems of localized spins coupled by
carriers, such as electrons bound to donors \cite{Bao}, magnetic
impurities \cite{Fer} and nuclear spins \cite{Kane}.

This work was supported by ONR, DARPA and ARDA/ARO. G.~R.~wishes
to thank A.~K.~Rajagopal for helpful discussions.


\begin{references}
\bibitem{Pier-note} Multiple scattering effects were accounted for
in ref.~\cite{Pier2}, where the problem of two localized spins
interacting via an itinerant exciton was solved utilizing
scattering matrix formalism. This approach goes beyond the second
order perturbation result given in Eq.~\eqref{J12} but
necessitates the use of a model separable potential.
\end{references}
\end{document}